\documentclass{article}

\usepackage{times}

\usepackage{url}
\usepackage{graphicx}        
\usepackage{subfigure}
\usepackage{amsmath}
\usepackage{amssymb}

\newtheorem{theorem}{Theorem}
\newtheorem{lemma}[theorem]{Lemma}

\newtheorem{definition}[theorem]{Definition}
\newenvironment{proof}[1][Proof]{\begin{trivlist}
\item[\hskip \labelsep {\bfseries #1}]}{\end{trivlist}}

\title{The Continuous Node Degree: a New Measure for Complex Networks}

\author{
Sherief Abdallah$^{1,2,\ast}$\\
\small{$^{1}$Faculty of Informatics, the British University in Dubai, UAE}\\
\small{$^{2}$School of Informatics, University of Edinburgh, UK}\\
\normalsize{$^\ast$To whom correspondence should be addressed; E-mail:  shario@ieee.org.}
}

\date{}

\begin{document}


\maketitle

\begin{abstract} 
A key measure that has been used extensively in analyzing complex networks is the degree of a node (the number of the node's neighbors). Because of its discrete nature, when the degree measure was used in analyzing weighted networks, weights were either ignored or thresholded in order to retain or disregard an edge. Therefore, despite its popularity, the degree measure fails to capture the disparity of interaction between a node and its neighbors.
 
We introduce in this paper a generalization of the degree measure that addresses this limitation: the continuous node degree (C-degree). We prove that in general the C-degree reflects how many neighbors are effectively being used (taking interaction disparity into account) and if a node interacts uniformly with its neighbors (no interaction disparity) the C-degree of the node becomes identical to the node's (discrete) degree. We analyze four real-world weighted networks using the new measure and show that the C-degree distribution follows the power-law, similar to the traditional degree distribution, but with steeper decline. We also show that the ratio between the C-degree and the (discrete) degree follows a pattern that is common in the four studied networks.
 \end{abstract}

\section{Introduction}
\label{sec-intro}

Network analysis is an interdisciplinary field of research that spans over biology, chemistry, computer science, sociology, and others. A key measurement that has been used extensively in analyzing networks is the \emph{degree} of a node. A node's degree is the number of edges incident to that node. Intuitively, the degree of a node reflects how connected the node is. This simple measure (along with other network measures) allowed the discovery of universal patterns in networks, such as the power law of the degree distribution \cite{barabasi99s,faloutsos99sigcomm}.

One of the limitations of the degree measure is that it ignores any disparity in the interaction between a node and its neighbors. In other words, the degree measure assumes uniform interaction across each node's neighbors. 
This can result in giving an incorrect perception of the \emph{effective} node degree. For example, a person may have 10 or more acquaintances but mainly interacts with only two of them (friends). Should that person be considered 2 times more connected than a person with only 5 acquaintances but also interacting primarily with two of them?

Several network measures were proposed to analyze weighted networks \cite{almaas04n,barth05pa,boccaletti06pr,mcglohon08kdd}, where an edge's weight quantifies the amount of interaction over the edge. However, none of the previously developed measures is a proper \emph{generalization} of the degree measure. A proper generalization of the degree measure that captures the disparity of interactions needs to satisfy three properties. The first property is preserving the maximum traditional degree: if all weights incident to a node  are equal (maximum utilization of neighbors), then the generalized degree is maximum and should be equal to the traditional (discrete) degree. The second property is preserving the minimum traditional degree: if all edges incident to a node have weights that are almost zero except one edge that has a weight much larger than zero (the node interacts primarily with one neighbor) then the generalized degree should be very close to 1. The third and final property is the consistent handling of disparity: the partial order imposed by the generalized degree on any two nodes needs to be consistent with the previous two properties. Intuitively, this means the more equal the weights are, the higher their generalized degree should be. We formalize these properties into axioms in the following section. 

A generalization of the degree measure is significant because it bridges the gap between the extensive research made using the degree (which ignored weights) and the research on weighted networks. Furthermore, it allows more accurate analysis of the networks that were previously analyzed using the degree measure. For example, it is known that the degree distribution of the Internet follows the power law \cite{faloutsos99sigcomm}. However, if one takes the disparity of interactions into account, does the \emph{effective} degree distribution of the Internet still follow a power law?

We introduce in this paper a new measure for analyzing weighted networks: the \emph{continuous degree (C-degree)}. What sets our measure apart from previous work is that it is a continuous generalization of the degree measure that captures the disparity of interaction. In particular, we prove that if every node interacts with all its neighbors equally, then the C-degree becomes identical to traditional (discrete) degree measure of the same node. However, if there is a disparity in a node's interaction with its neighbors, then the C-degree will capture such disparity, unlike the traditional degree measure.

An implicit assumption for using our measure is that network weights quantify the amount of interaction, therefore all the weights are positive.
We analyze four real world networks using the new measure and show that the C-degree distribution still follows the power law (but with a steeper power coefficient than the discrete degree distribution). We also show that the ratio between the C-degree and the traditional degree is bounded.

\section{Background}
\label{sec-back}

A network (graph) is defined as $N=\langle V,E \rangle$, where $V$ is the set of network nodes (vertices) and $E$ is the set of edges (links) connecting these nodes. 
The degree of a node $v \in V$ is $k(v)=|E(v)|$, where $|E(v)|$ is the number of edges incident to node $v$. The degree distribution $P(k)$ measures the frequency of a particular degree $k$ in a network: $P(k=u) = |\{v:v \in V \wedge k(v)=u\}|$. The degree distribution is a common method for combining the degrees of all network nodes into one measure that summarizes and characterizes the network.

This paper focuses on weighted networks where the weight of an edge $w(e) \ge \epsilon$ quantifies the amount of interaction across the edge $e$ and $\epsilon$ is a small constant greater than (and close to) zero. We call networks that satisfy this property \emph{interaction networks}. For example, an edge weight can represent the number of times a person calls a friend or the number of packets transmitted on an Internet link. For convenience, we define for each node $i$ the set of incident weights $W(i)=\{w(e) : e \in E(i)\}$. 
A node strength $s(i)= \sum_{w \in W(i)} w$ \cite{boccaletti06pr} is the summation of weights incident to a node $v$. 

Before designing a measure that generalizes the traditional node degree so that it captures the disparity of interactions among neighbors, one needs to clearly define properties that this new measure needs to satisfy. Let $r(i)$ be a generalized degree of node $i$. The first property of $r(i)$ is preserving the maximum traditional degree: if all weights incident to a node $i$ are equal, then the generalized degree of node $i$, $r(i)$, is maximum and equal to the traditional (discrete) degree of node $i$, i.e $r(i)=k(i)$. The second property is preserving the minimum traditional degree: if all edges incident to node $i$ have weights that are almost zero except one edge that has weight much larger than zero (i.e. node $i$ interacts primarily with one neighbor) then the generalized degree of node $i$ should be very close to $1$. 

The third and final property is the consistent handling of disparity: the partial order imposed by the generalized measure on any two nodes need to be consistent with the previous two properties. Intuitively, this means the more equal the weights are, the higher their generalized degree should be. 
Many partial ordering are possible, but here we define what we believe is the minimum requirement.
If two nodes $i$ and $j$ have the same number of neighbors $n$, the same strength, and have $n-2$ common weights, then the generalized degree should be inversely proportional to difference between the uncommon weights of each node. For example, suppose $W(v1) = \{5,5,5,5\}, W(v2)=\{9,5,5,1\}, W(v3)=\{9,8,2,1\}$ and $W(v4)=\{20-3\epsilon,\epsilon,\epsilon,\epsilon\}$. The third property then require the generalized degree measure $r$ to impose the following ordering: $r(v1)>r(v2)>r(v3)$. All three properties require that $k(v1)=r(v1)>r(v2)>r(v3)>r(v4)\approx 1$.
The following axioms formalize the three properties that a generalized degree measure $r$ needs to satisfy. 

\begin{enumerate}
\item \textbf{Preserving maximum degree:} $r(i)=k(i)=|E(i)|$ iff $W(i)=W_{max}(i)$, where $W_{max}(i)=\{w_l: w_l=\frac{s(i)}{k(i)} \and 1 \le l \le k(i)\}$.  
\item \textbf{Preserving minimum degree:} $r(i)$ is close to 1 iff $\exists u: w(u) >> \epsilon \and \forall v \ne u: w(v)= \epsilon$. 
\item \textbf{Consistent disparity:} $\forall i,j$ such that $k(i)=k(j)=n, s(i)=s(j), |W(i) \bigcap W(j)|=n-2, \{w_{i1},w_{i2}\} = W(i) - W(j),$ and $\{w_{j1},w_{j2}\} = W(j) - W(i)$: if $|w_{i1}-w_{i2}| < |w_{j1}-w_{j2}|$ then $r(i) > r(j)$.
\end{enumerate}

Some of the previous work used a cutoff weight-threshold in order to either include or exclude a weighted edge and then computed the degree distribution normally \cite{chapanond05cmot,kalisky06phr}. Such an approach, however, does not properly handle the disparity of interaction among neighbors, but rather approximates a weighted network with an unweighted network.

Surveying all network measures that were proposed to analyze weighted networks is beyond the scope of this paper. Instead, we focus on a sample of these measures that are mostly related to our contribution (interested reader may refer to survey papers on the subject such as \cite{boccaletti06pr}). The \emph{weight distribution} $P(w)$ is similar to the degree distribution except that it measures the frequency of a particular edge weight. This measure neither generalizes the degree distribution nor does it capture the disparity in interaction between an individual node and its neighbors.

The strength of a node becomes identical to the node's degree if all weights are equal to 1. The strength measure, however, fails to capture the disparity of interaction between an individual node and its neighbors (the consistency axiom). 
A more recent work \cite{mcglohon08kdd} analyzed a graph's total weight, $\sum_{e\in E} w(e)$, against the graph's total number of edges, $|E|$, over time. That work also analyzed the degree of a node, $k(v)$, against the node's strength, $s(v)$. These measures again fail to capture the disparity in interaction between a node and its neighbors.

The network measure $Y(v)=\sum_{e \in E(v)} \left(\frac{w(e)}{s(v)}\right)^2$ successfully captures the disparity of interaction within a node $v$ \cite{almaas04n}.
However, the $Y$ measure is not a \emph{generalization} of the degree measure as it fails to satisfy the first two axioms. 

An interesting method for generalizing unweighted network measures (including the node degree) to weighted networks is generating an ensemble of unweighted networks that are sampled from the original weighted network \cite{ahnert07pre}. The underlying assumption the method is that the weight of an edge reflects the probability of generating the edge in a sample network. The effective node degree is then the average over the samples. While the ensemble approach satisfy the first two axioms, it fails to satisfy the third axiom, the consistency in handling disparity. 
The following section presents our proposed measure, the continuous degree, and proves that it satisfies all the three axioms of a generalized node degree.

\section{The Continuous Degree, C-degree}
\label{sec-ied}

The inherent problem with the degree is it being \emph{discrete}. A neighbor is either counted in the degree or not. We propose a generalization of the degree measure that takes edge weights into account. 

\begin{definition}
The \emph{C-degree} of a node $v$ in a network is r(v), where
\[ 
r(v) = \left\{ \begin{array}{ll}
			0 & \textrm{if $v$ is disconnected} \\
			2^{\left(\sum_{e\in E(v)} \frac{w(e)}{s(v)} \log_2 \frac{s(v)}{w(e)}\right)} & otherwise
			\end{array} \right.
\]
\end{definition}

Where $s(v)=\sum_{e\in E(v)} w(e)$ is the strength of node $v$. Intuitively, the quantity $\frac{w(e)}{s(v)}$ represents the probability of an interaction over an edge $e$. The set $\left\{\frac{w(e)}{s(v)}: e \in E(v)\right\}$ is the interaction probability distribution for node $v$. The quantity $\sum_{e\in E(v)}\left[ \frac{w(e)}{s(v)} \log_2 \frac{s(v)}{w(e)} \right]$ is then the entropy of the interaction probability distribution, or how many bits are needed to encode the interaction probability distribution. The entropy quantifies the disparity in the interaction distribution: the more uniform the interaction distribution is, the higher the entropy and vice versa.\footnote{This is in contrast to the Y measure, which decreases if the interaction distribution becomes more uniform.} The purpose of the power 2 is to convert the entropy back to the number of neighbors that are effectively being used. 

\begin{figure}[htbp] \centering
\includegraphics[width=3.5in]{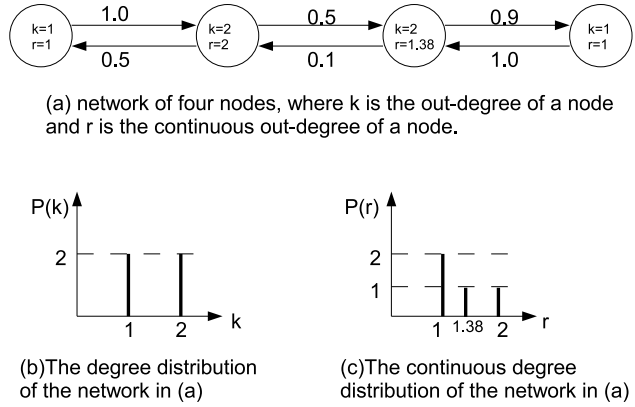}
      \caption{\small Continuous vs discrete degree distributions.}
      \label{fig-cdd}
    \end{figure}

Figure \ref{fig-cdd} compares the continuous degree distribution to the (discrete) degree distribution in a simple weighted network of four nodes. A node on the boundary has an out degree of 1, while an internal node has an out degree of 2. Intuitively, however, only  one of the internal nodes is fully utilizing its degree of 2 (the one to the left), while the other node (to the right) is mostly using one neighbor only. The C-degree measure captures this and shows that only one internal node has a C-degree of 2 while the other internal node has a C-degree of 1.38. In the remainder of this section we prove that the C-degree satisfies all the three properties defined earlier.

In the remainder of this section we prove that the C-degree satisfies all the three properties defined earlier.

\begin{theorem}
\label{thrm-cd}
The C-degree measure satisfies all the three axioms of a generalized node degree.
\end{theorem}
\begin{proof}
The proof directly follows from the following three lemmas.
\end{proof}

\begin{lemma}
The C-degree satisfies the minimum degree axiom.
\end{lemma}

\begin{proof}
When all weights are close to zero except only one weight that is much bigger than zero, then the entropy (the exponent of the C-degree) is close to zero, and therefore the C-degree is close to 1.
\end{proof}

\begin{lemma}
\label{lm-cdmax}
The C-degree satisfies the maximum degree axiom.
\end{lemma}

\begin{proof}
Under uniform interaction, all the weights incident to a node $v$ are equal to a constant $W_v$. Therefore 
\[
\forall v \in V, e \in E(v): \frac{w(e)}{s(v)}=\frac{W_v}{\sum_{e\in E(v)}W_v}=\frac{1}{k(v)}
\] 
We then have
\begin{align*}
\forall v: r(v) & = 2^{\sum_{e \in E(v)} \frac{w(e)}{s(v)} \log_2 \frac{s(v)}{w(e)}} \\
& = 2^{\sum_{k(v)} \frac{1}{k(v)} \log_2(k(v))} \\
& = 2^{\log_2(k(v))} \\
& = k(v)
\end{align*}

In other words, both the degree and the C-degree of a node become equivalent under uniform interaction. The C-degree is also maximum in this case, because the exponent is the entropy of the interaction distribution, which is maximum when the interaction is uniform over edges. 
\end{proof}

\begin{lemma}
The C-degree satisfies the consistent disparity property.
\end{lemma}

\begin{proof}
Let $i,j$ be two nodes such that $k(i)=k(j)=n, s(i)=s(j)=s, |W(i) \bigcap W(j)|=n-2, \{w_{i1},w_{i2}\} = W(i) - W(j),\{w_{j1},w_{j2}\} = W(j) - W(i)$, and $|w_{i1}-w_{i2}| < |w_{j1}-w_{j2}|$. Without loss of generality, we can assume that $w_{i1} \ge w_{i2}$ and $w_{j1} \ge w_{j2}$, therefore $w_{i1}-w_{i2} < w_{j1}-w_{j2}$.  We also have 
\[
\frac{w_{i1}+w_{i2}}{s}=1 - \sum_{w \in W(i) \bigcap W(j)}\frac{w}{s} = \frac{w_{j1}+w_{j2}}{s} = c
\], therefore 
\[
\frac{w_{j1}}{s} > \frac{w_{i1}}{s} \ge \frac{c}{2} \ge c-\frac{w_{i1}}{s} > c-\frac{w_{j1}}{s}
\] 
Then from Lemma \ref{lemma-ent}
\begin{align*}
h(c,\frac{w_{i1}}{s}) > h(c,\frac{w_{j1}}{s}) \\
-\frac{w_{i1}}{s}lg(\frac{w_{i1}}{s}) - (c-\frac{w_{i1}}{s})lg(c-\frac{w_{i1}}{s}) > -\frac{w_{j1}}{s}lg(\frac{w_{j1}}{s}) - (c-\frac{w_{j1}}{s})lg(c-\frac{w_{j1}}{s})
\end{align*}
Therefore $H(i) > H(j)$, because the rest of the entropy terms (corresponding to $W(i) \bigcap W(j)$ are equal, and consequently $r(i)>r(j)$.
\end{proof}

\begin{lemma}
\label{lemma-ent}
The quantity $h(c,x)=-xlg(x) - (c-x) lg(c-x)$ is symmetric around and maximized at $x=\frac{c}{2}$ for $c\ge x > 0$.
\end{lemma}
\begin{proof}
Symmetry: $h(c,\frac{c}{2}+\delta)=-(\frac{c}{2}+\delta)lg(\frac{c}{2}+\delta) - (\frac{c}{2}-\delta) lg (\frac{c}{2}-\delta) = h(c,\frac{c}{2}-\delta)$. 
Maximum: $h(c,x)$ is maximized when $\frac{\partial h(c,x)}{\partial x} = 0 = -1 -lgx + 1 + lg(c-x)$, therefore $x=c-x = \frac{c}{2}$.
\end{proof}

Theorem \ref{thrm-cd} means that, in the simple case of uniformly distributed utilization of connections, the continuous degree distribution \emph{preserves} well-known properties of the discrete degree distribution, such as the power law. It remains, however, to investigate what laws the continuous degree distribution follows in the more general setting of interaction networks. In the following section we use the C-degree measure to analyze four real-world weighted networks.

\section{Case Studies}
\label{sec-results}

We have analyzed four real world weighted networks \footnote{Available through http://www-personal.umich.edu/~mejn/netdata/}
that capture coauthorships between scientists. Three of which were extracted from preprints on the E-Print Archive \cite{newman01pnas}: condensed matter (updated version of the original dataset that includes data between Jan 1, 1995 and March 31, 2005), astrophysics, and high-energy theory. The fourth network represents coauthorship of scientists in network theory and experiment \cite{newman06phr}.

\begin{figure}[htbp]
  \subfigure[condensed matter]{
		\includegraphics[width=2.5in]{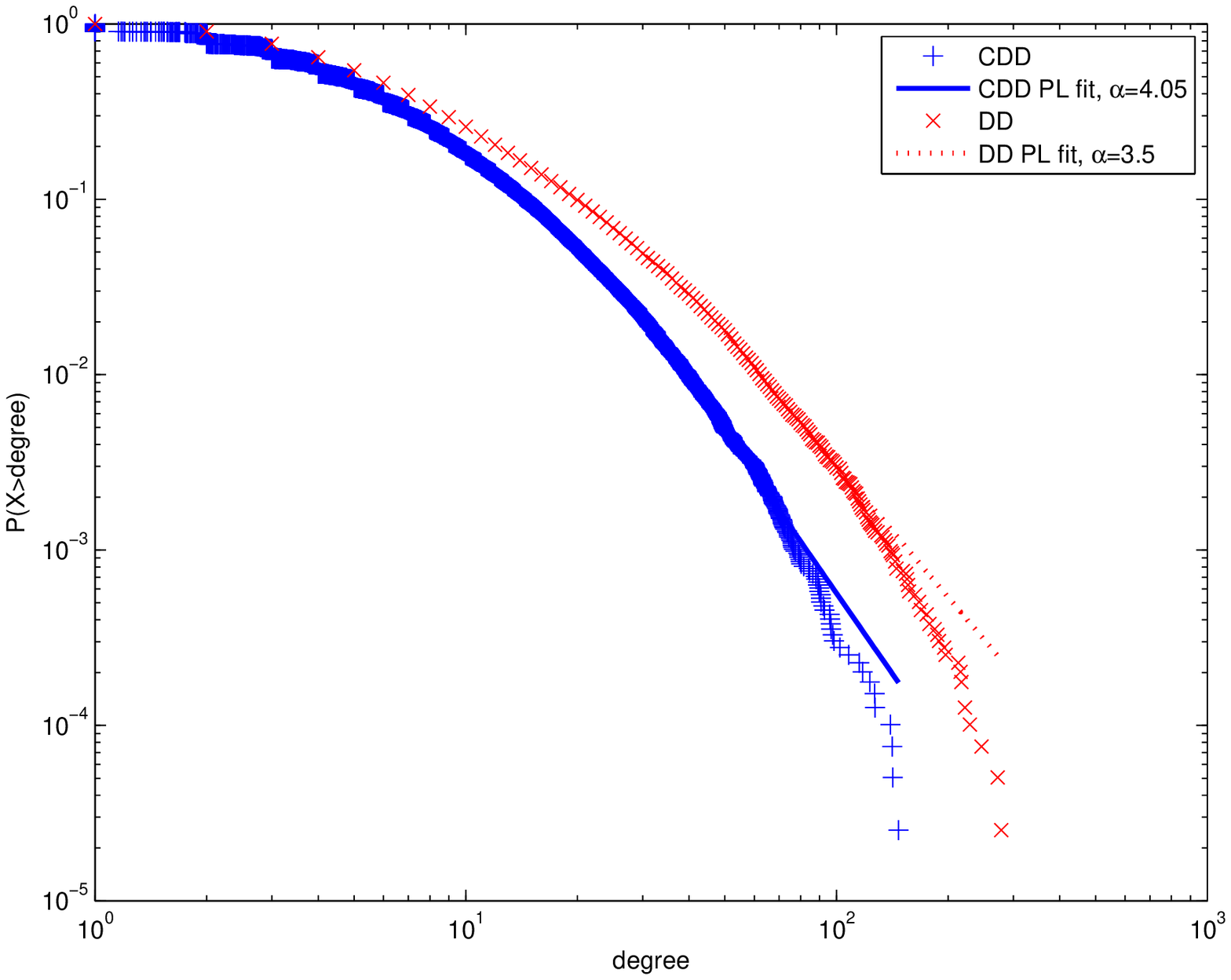} 
}
  \subfigure[astrophysics]{
		\includegraphics[width=2.5in]{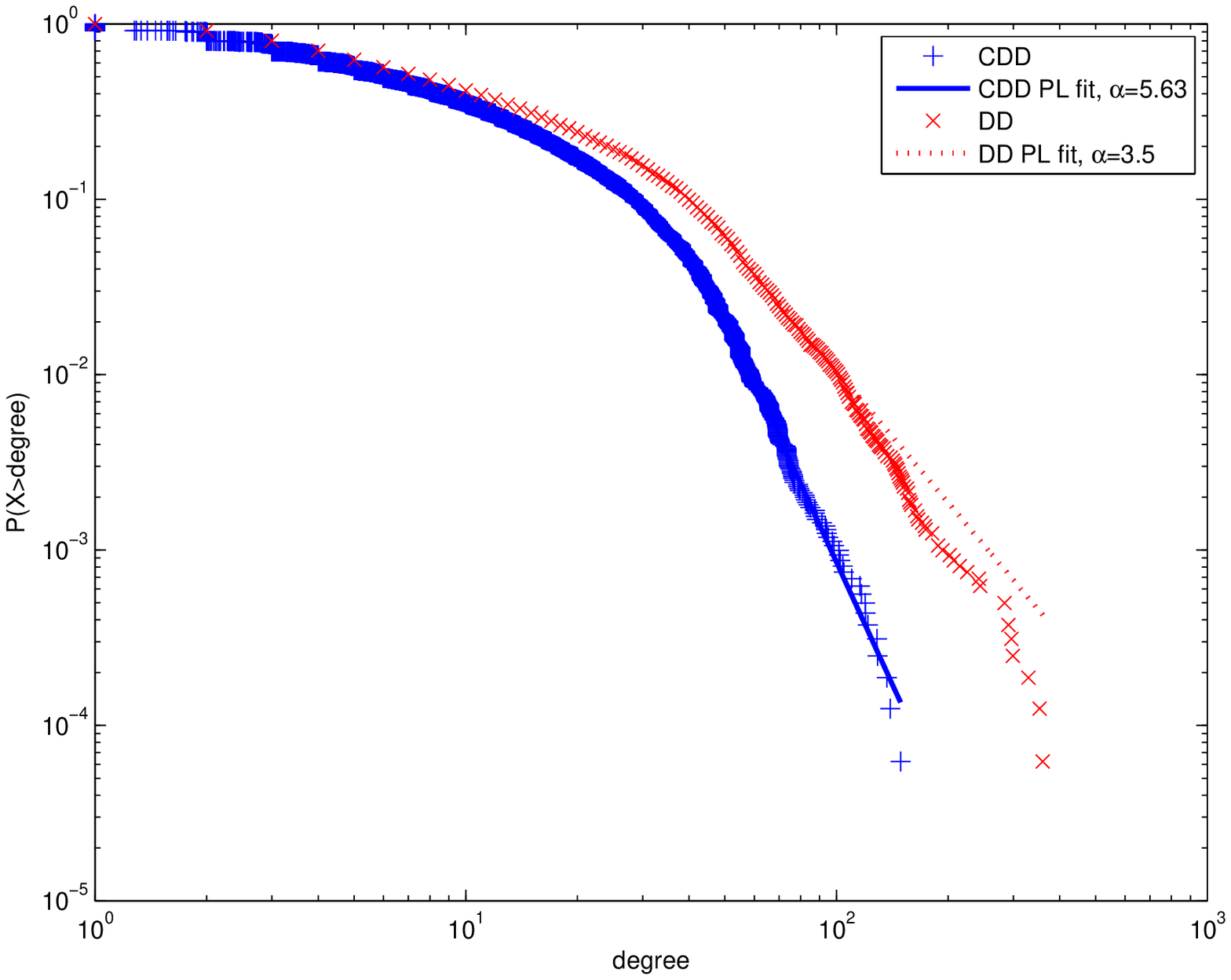} 
}
  \subfigure[network theory]{
		\includegraphics[width=2.5in]{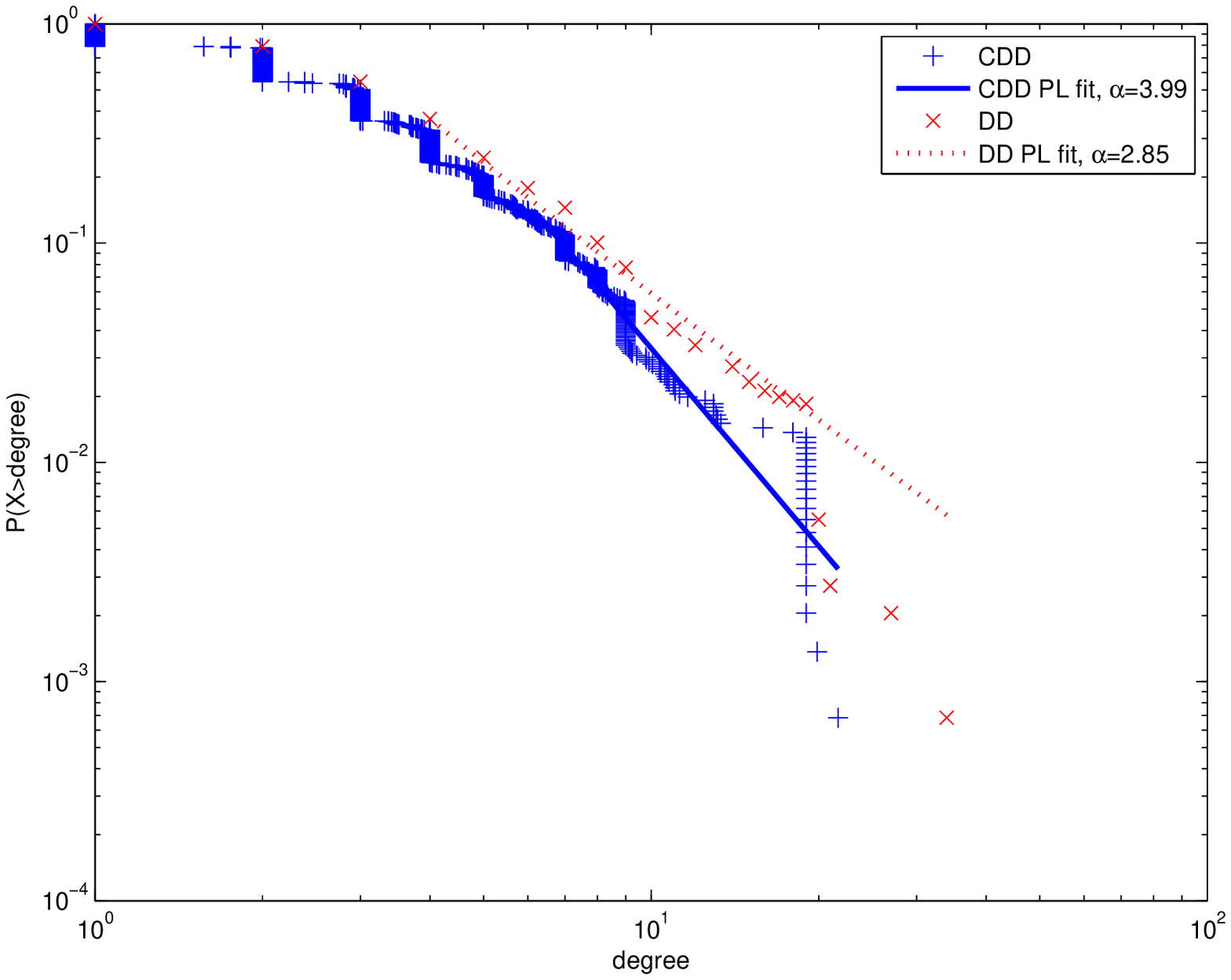} 
}
  \subfigure[high-energy theory]{
		\includegraphics[width=2.5in]{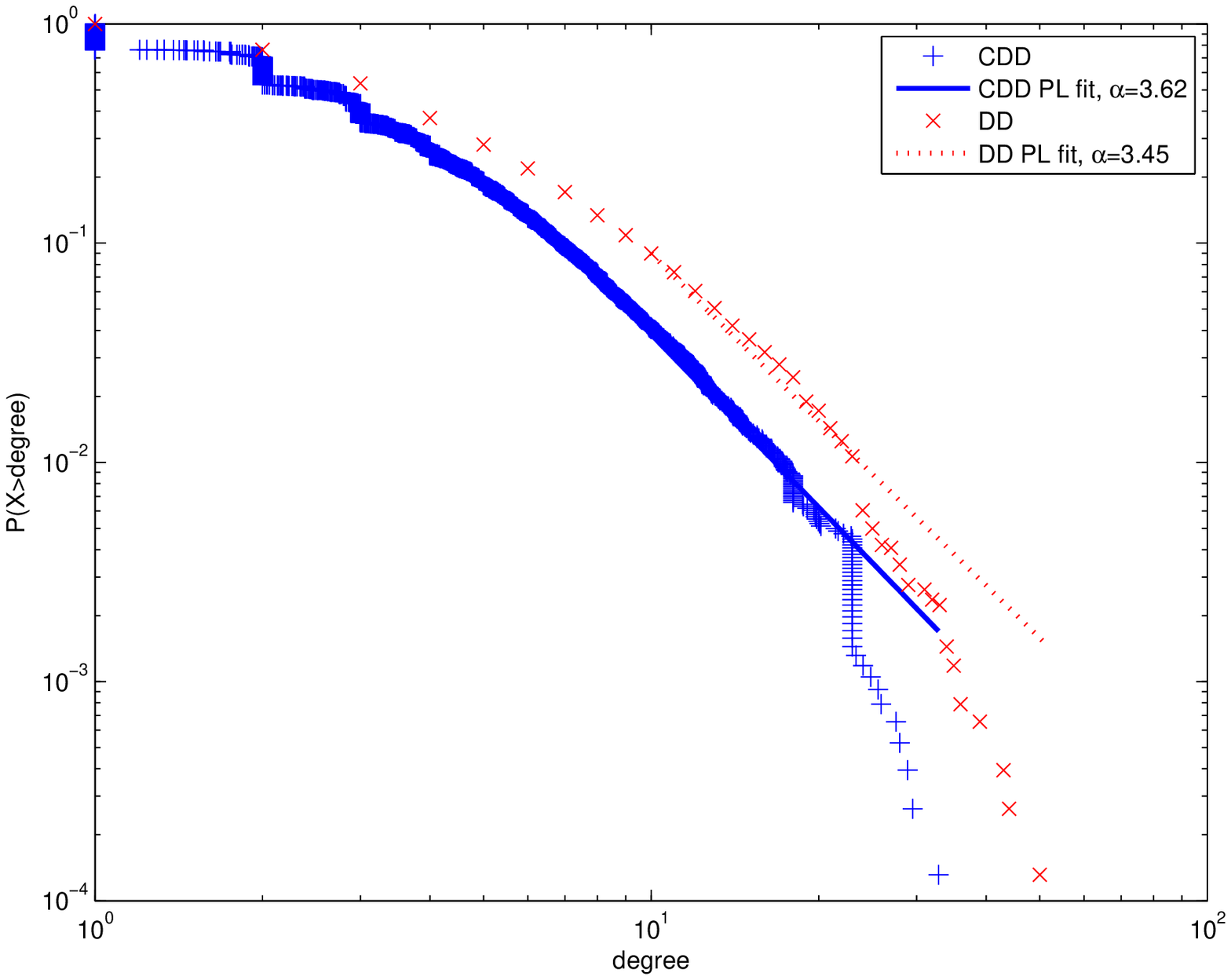} 
}
\caption{{\small Comparing the DD with CDD for the collaboration networks.}}
      \label{fig-power}
\end{figure}

It was shown that the degree distribution of many real networks follows the power law \cite{barabasi99s,boccaletti06pr}. A degree distribution follows the \emph{power law} if $P(k) \propto k^{-\gamma}$, where $\gamma$ is a constant. 
Figure \ref{fig-power} displays the C-degree distribution (CDD) and the (discrete) degree distribution (DD) for the four collaboration network. The figure uses log-log scale with the power law fit (based on \cite{clauset07arxiv}\footnote{Source code available from http://www.santafe.edu/\~aaronc/powerlaws/}). The CDD preserves the power-law behavior, but with steeper decline, which is consistent with Lemma \ref{lm-cdmax}.

\begin{figure}[htbp]
  \subfigure[condensed matter]{
		\includegraphics[width=2.5in]{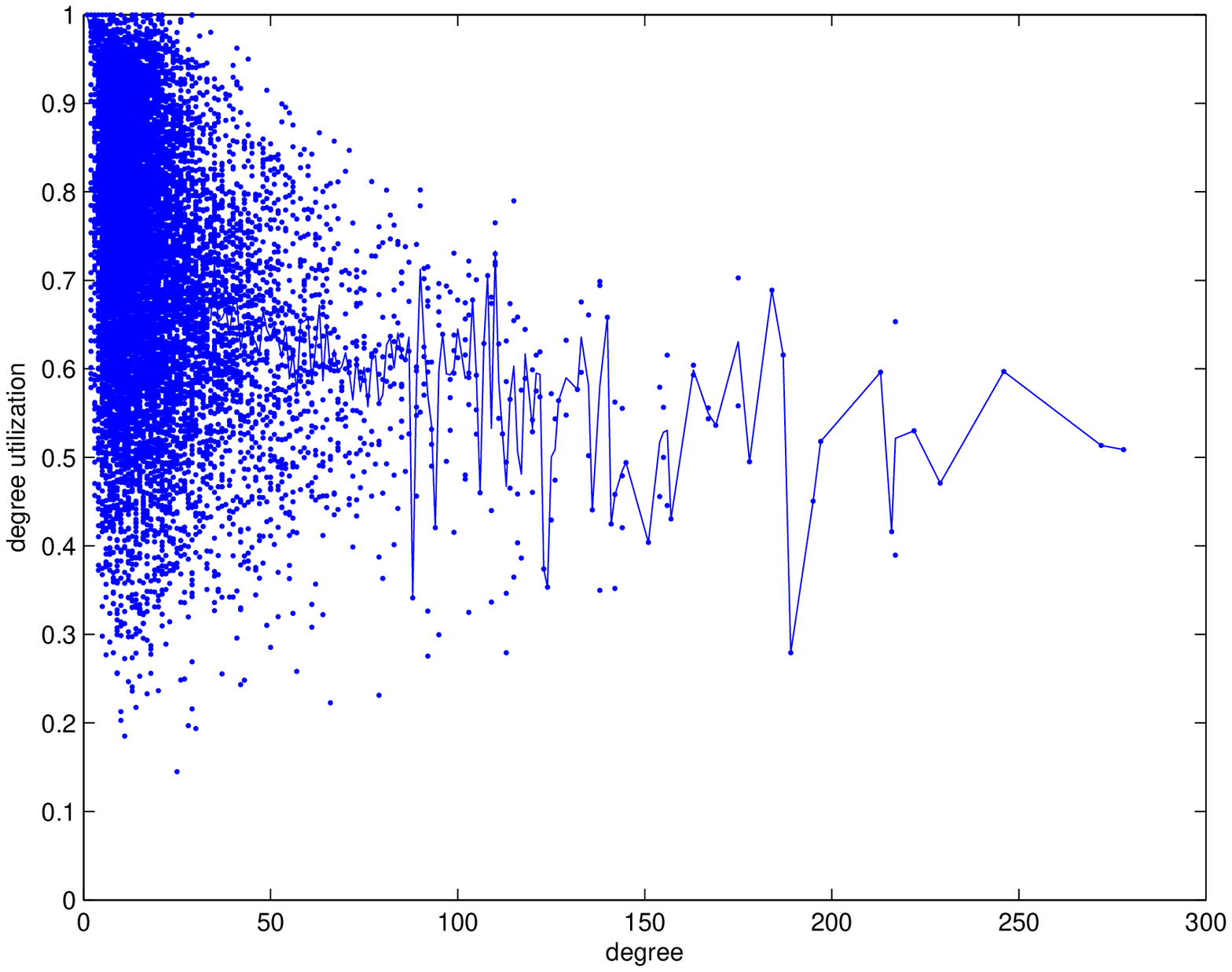} 
}
  \subfigure[astrophysics]{
		\includegraphics[width=2.5in]{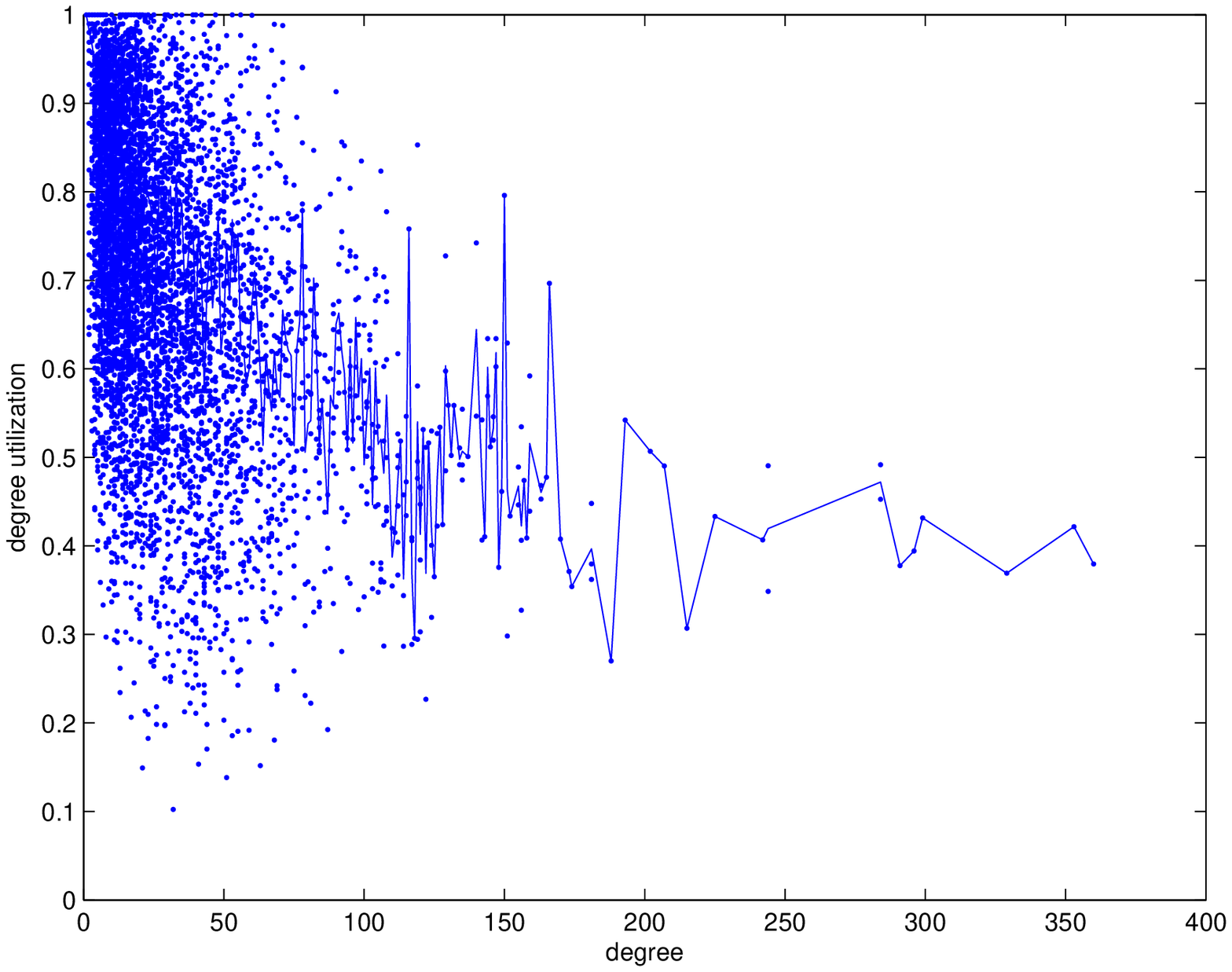} 
}
  \subfigure[network theory]{
		\includegraphics[width=2.5in]{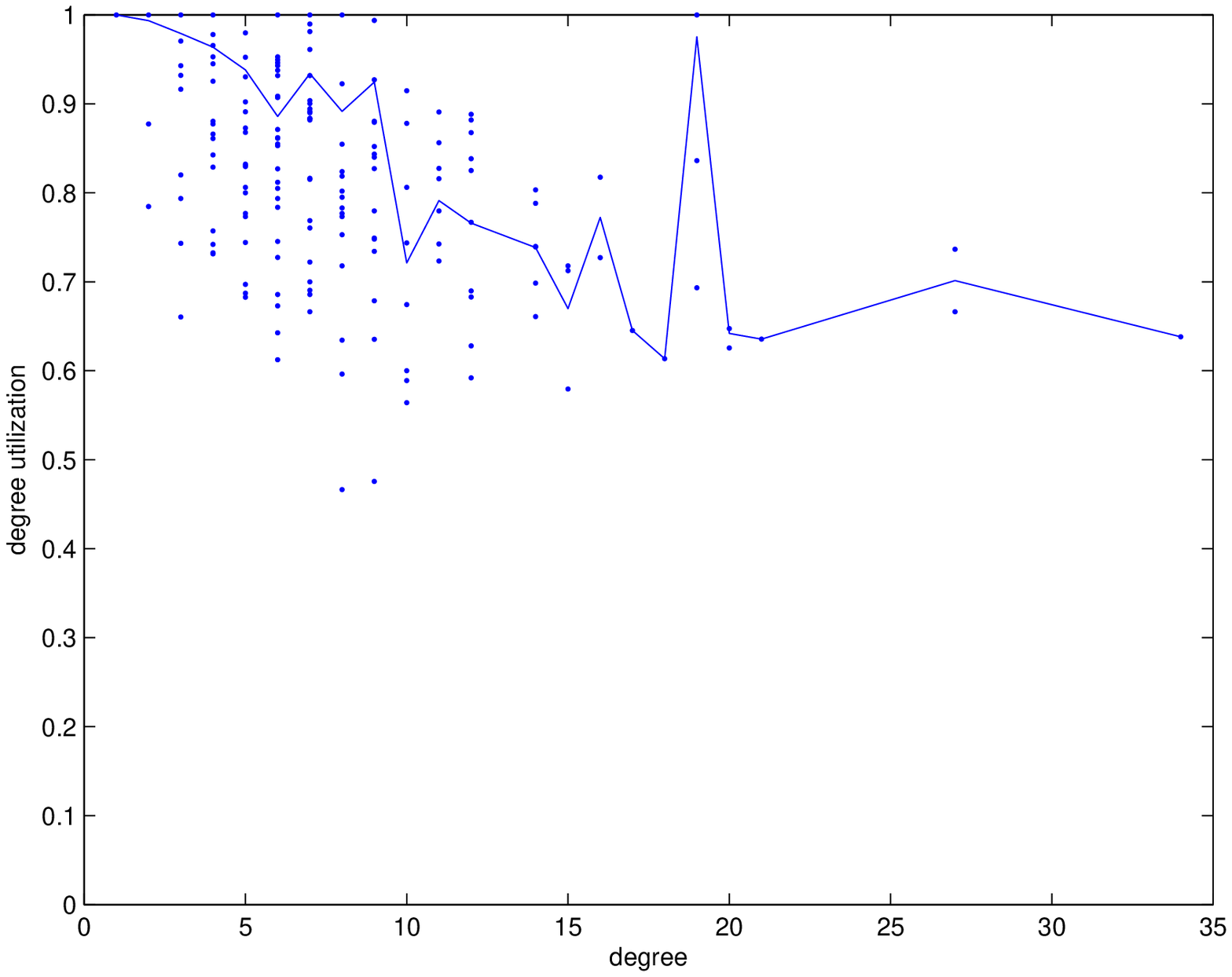} 
}
  \subfigure[high-energy theory]{
		\includegraphics[width=2.5in]{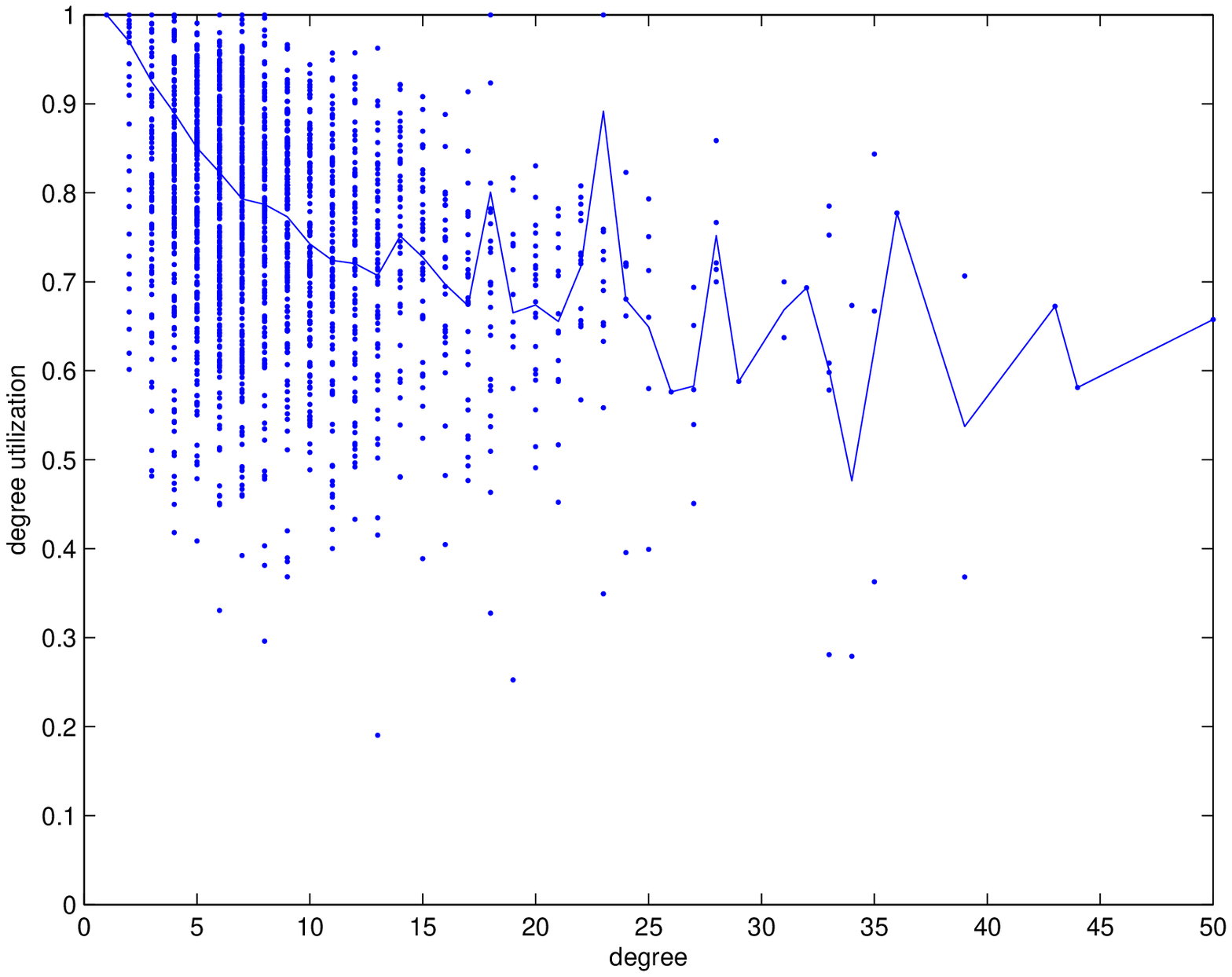} 
}
\caption{{\small Scatter plot of a node degree against its degree utilization for the four collaboration networks. the average utilization per degree is also plotted.}}
      \label{fig-cdpd}
\end{figure}

One would expect that as the degree of a node increases, the node will interact primarily with a smaller subset of neighbors. To verify this intuition, we define the \emph{degree utilization} metric as the ratio between the C-degree and the degree of a node: $u(v)=\frac{r(v)}{k(v)}$. The degree utilization metric captures the percentage of links that a node uses effectively, therefore we expect the degree utilization to decrease as the degree increases. Figure \ref{fig-cdpd} plots the degree utilization against the (discrete) degree for the four collaboration networks. A common pattern emerges in the four networks. For low degrees, the degree utilization is relatively high (a node with few links makes the best of them).  
For node degree greater than some constant the bias towards high degree utilization disappears. However, and to our surprise, 
a \emph{cone} is observed, which starts wide at low degrees and gets narrower as the degree increases (the average degree utilization is plotted as a line in the figure).
    
\section{Conclusion}
\label{sec-conclude}
We introduced in this paper a new measure for analyzing weighted networks: the C-degree. We proved that our measure is a continuous generalization of the discrete degree, and therefore bridges the gap between the analysis using the degree in unweighted networks and weighted networks, where weights quantify interaction among nodes. We demonstrated the applicability of this new measure by analyzing four real-world weighted networks. We showed that the C-degree distribution follows the power law, but with a steeper power-coefficient. We also investigated the ratio between the C-degree and the traditional degree and showed that the on average it is lower bounded, even for nodes with high-degree.

\bibliographystyle{abbrv}
\bibliography{references}

\end{document}